# Observation of nonreciprocal transverse localization of light

Shun Liang[1], Changchang Li[1], Wenqing Yu[1], Zhenzhi Liu[1], Changbiao Li[1], Yanpeng Zhang[1], Guillaume Malpuech[2], Dmitry Solnyshkov[2,3,†], Hui Jing[4,*], Zhaoyang Zhang[1,‡]

[1]*Key Laboratory for Physical Electronics and Devices of the Ministry of Education & Shaanxi Key Lab of Information Photonic Technique, School of Electronic and Information Engineering, Faculty of Electronic and Information Engineering, Xi'an Jiaotong University, Xi'an, 710049, China*

[2]*Institut Pascal, PHOTON-N2, Université Clermont Auvergne, CNRS, Clermont INP, F-63000 Clermont-Ferrand, France*

[3]*Institut Universitaire de France (IUF), 75231 Paris, France*

[4]*Key Laboratory of Low-Dimensional Quantum Structures and Quantum Control of Ministry of Education, Department of Physics and Synergetic Innovation Center for Quantum Effects and Applications, Hunan Normal University, Changsha 410081, China*

Emails: †dmitry.solnyshkov@uca.fr, *jinghui@hunnu.edu.cn, ‡zhyzhang@xjtu.edu.cn

Magnetic-free nonreciprocal optical devices that can prevent backscattering of signals are essential for integrated optical information processing. The achieved nonreciprocal behaviors mostly rely on various dispersive effects in optical media, which give rise to dispersive modulations of the transverse beam profile, such as spatial broadening and discretization, of the incident signals. Such deformation inevitably reduces the matching with subsequent components for information processing. Here we experimentally demonstrate the nonreciprocal transverse localization of light in a moiré photonic lattice induced in atomic vapors. When the probe field is set to co- or counter-propagate with the coupling field formed by superposing two identical honeycomb beams in a certain rotation angle, the output pattern can exhibit localized or dispersive behavior. The localization in the forward case is derived from the moiré structure, and the nonreciprocal behaviors (in both beam size and transmitted intensity) are introduced by the thermal motion of atoms. The thermal-motion-induced Doppler effect can destroy the coherent condition for electromagnetically induced transparency in the backward case, because of which the probe beam becomes immune to the modulation of the coupling field. The current work provides an approach to control the transverse beam profile in one-way transmission.

The development of integrated optical information processing techniques promotes the quest for magnetic-free nonreciprocal optical devices, which usually break the time-reversal symmetry to earn the desired capability in enforcing the unidirectional transmission of optical waves [1-4]. Up to now, the design of high-performance nonreciprocal optical functionalities has been realized with plenty of approaches without the presence of bulky magnets but bypassing Lorentz reciprocity, including valley polarization pumping [5], optical nonlinearity [6-9], Autler-Townes splitting [10], Sagnac effect in resonators [11, 12], chiral light-matter interactions [13, 14], optically-induced magnetization [15] and Brillouin scattering [16], to name a few. In the meanwhile, optical nonreciprocity has also been proposed under the frame work of obeying time-reversal symmetry [17, 18]. The implementations of magnet-free optical nonreciprocity, irrespective of satisfying time-reversal symmetry or not, mostly lie in homogeneous bulk media [19, 20], optical resonators [14, 21-24] and spatially discrete photonic structures [25-30]. In most adopted nonreciprocal optical platforms, the behaviors of light are essentially governed by the dispersion of optical media, which can give rise to dispersive modulations (such as spatial broadening and discretization) on the incident light, yet with high nonreciprocal contrast on transmitted intensities. Such deforming effects from nonreciprocal optical devices can reduce the matching between the signals and subsequent components for information processing and transmission.

In 2018, the concept of nonreciprocal localization of photons was proposed in a 1D moving photonic lattice with an embedded static defect [31]. Such a dynamical periodic configuration based on electromagnetically induced transparency (EIT) [32] in an atomic system produces a localized state or scattering mode for photons travelling in opposite directions, leading to nonreciprocal transmission. Very recently, nonreciprocal 1D solitons were reported in a spinning Kerr resonator [33] and in active metamaterials [34]. Optical solitons, resulting from the balance between dispersion and nonlinearity [35], can also cancel the dispersive effects but with a threshold for the input power. This indicates that nonreciprocal solitons are applicable for situations involving relatively strong signals. However, no transverse effects were studied in these works. The transverse beam profile is important for beam matching between the non-reciprocal devices and follow-up optical components. Transverse localization could be promising in counteracting the dispersive propagating effects in magnetic-free nonreciprocal optical devices, but it has not been experimentally observed so far.

In this work, by combining the Doppler frequency shift from the thermal motion of atoms and flat bands from moiré structures, we demonstrate the nonreciprocal optical transverse localization in a honeycomb moiré photonic lattice optically induced in a three-level Rb atomic vapor cell. Actually, thermal atoms with Doppler effect, which can produce direction-dependent frequency shift on involved laser fields to destroy the EIT effect, have acted as a powerful platform in conducting nonreciprocal propagation of light [36, 37] in the longitudinal $z$ direction. The moiré photonic lattice [38-40], characterized by flat bands with considerably suppressed transverse dispersion [41-44], is established by superposing two honeycomb photonic lattices with a twist angle of 27.8° in the $x$-$y$ plane. The required honeycomb lattices are "written" under the condition of EIT [45-48] by two identical honeycomb coupling beams from a spatial light modulator (SLM). This is also the first experimental realization of instantaneously reconfigurable moiré photonic lattice with the assistance of EIT, which has been exploited to theoretically design various moiré structures [49, 50]. By selectively launching the probe beam in the same (forward case) or opposite (backward case) direction with the coupling field, the output can exhibit localized or dispersive feature, due to the presence or not of EIT. The imbalanced output probe intensities in the forward and backward cases are also achieved, since EIT can effectively suppress the resonant absorption, as in the non-Hermitian skin effect [51, 52]. With two probe beams simultaneously injected into the moiré lattice in opposite directions, the obtained maximum isolation ratio is about 20.1 dB.

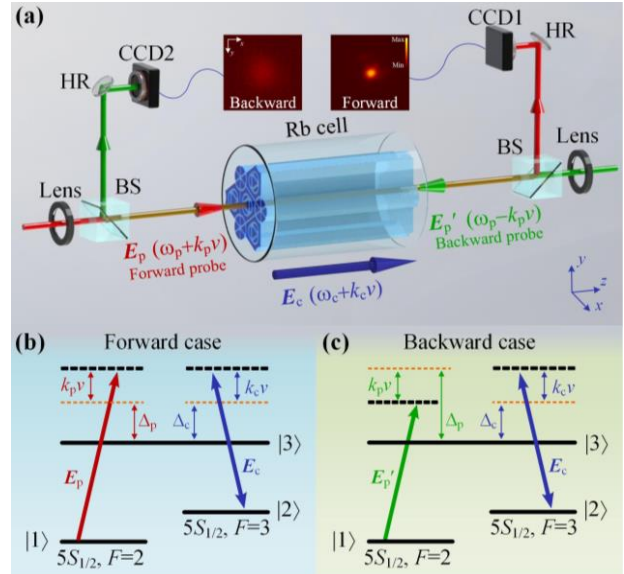

Fig. 1. (a) Experimental scheme. A SLM is employed to modulate the vertically polarized coupling field $E_c$ into a moiré pattern consisting of two honeycomb substructures. Two horizontally polarized probe beams $E_p$ (forward case) and $E_p'$ (backward case) are sent into the Rb atomic vapor cell in opposite directions, and their output patterns passing through the atomic medium are imaged onto two cameras (CCD1 and CCD2). The Λ-type energy-level atomic

configurations for forward (b) and backward (c) cases with Doppler effect considered.

Figure 1(a) depicts the experimental principle. A coupling field $\boldsymbol{E}_c$ (frequency $\omega_c$) with a transverse moiré intensity profile is established by superposing two identical honeycomb light beams generated from a phase-type SLM [43, 53]. The twist angle θ between the two honeycomb substructures can be controlled by loading different desired holographs, which possess the phase information of the targeted moiré structure (the detailed experimental setup and holography are provided in Fig. S1 in Supplemental Material). Then the weak forward probe beam $\boldsymbol{E}_p$ ($\omega_p$) is focused onto one site of the moiré pattern at normal incidence, and a three-level Λ-type energy-level structure is established inside atomic vapors.

The three-level atomic configuration for EIT contains two hyperfine ground states $F=2$ (level $|1\rangle$) and $F=3$ ($|2\rangle$) of $5S_{1/2}$ and an excited state $5P_{1/2}$ ($|3\rangle$), as given in Figs. 1(b) and 1(c). The field $\boldsymbol{E}_p$ ($\boldsymbol{E}_c$) drives the transition from the ground state $|1\rangle$ ($|2\rangle$) to the excited state $|3\rangle$ with detuning $\Delta_p$ ($\Delta_c$), which is denoted by the difference between the probe (coupling) frequency $\omega_p$ ($\omega_c$) and the natural frequency between the two energy levels it connects. For atomic vapors, the inevitable random thermal motion of the atoms can bring Doppler frequency shift for incident laser fields [32, 36]. The frequency shift is described by $k_i v$, with $k_i$ being the wave vector of the corresponding beam and $v$ being velocity of the atoms moving towards the beam. The amplitude of $k_i v$ is positively related to the atomic temperature and usually of the order of hundreds of megahertz. In the forward case [Fig. 1(b)] of $\boldsymbol{E}_p$ co-propagating with $\boldsymbol{E}_c$, $\Delta_p$ and $\Delta_c$ become $\Delta_p+k_p v$ and $\Delta_c+k_c v$, respectively. Considering the very minor difference in wavelengths of the probe and coupling beams, we have $k_p=k_p'\approx k_c$ and the two-photon detuning is $(\Delta_p+k_p v)-(\Delta_c+k_c v)\approx\Delta_p-\Delta_c$. This indicates that the influence of Doppler frequency shift is cancelled in the co-propagating case, which is also viewed as the Doppler-free situation. Under the EIT condition, the output probe can exhibit varying transverse distributions and localization after travelling through the moiré photonic lattices with different twist angles, when the two-photon detuning is set around $\Delta_p-\Delta_c=0$ for exciting the EIT effect. The transverse localization appears due to the peculiar profiles of potential and the band structures characterized by suppressed dispersions in moiré photonic structures [40].

In contrast, in the backward case with probe field $\boldsymbol{E}_p'$ (has the same parameters as $\boldsymbol{E}_p$ but is sent into the vapor cell oppositely) turned on, the two-photon detuning can be expressed as $(\Delta_p-k_p' v)-(\Delta_c+k_c v)\approx\Delta_p-\Delta_c-2k_p v$, see Fig. 1(c). It means that the Doppler shift destroys the two-photon resonant condition for EIT [32]. Thus, the backward probe beam $E_p'$ is immune to the modulation from the coupling field and experiences a strong resonant absorption together with a usual beam broadening, resulting in a weak and broad Gaussian pattern after dispersive propagation. As a consequence, the output characteristics for the forward and backward probe beams are nonreciprocal in both intensity and transverse profile, see the upper two insets in Fig. 1(a).

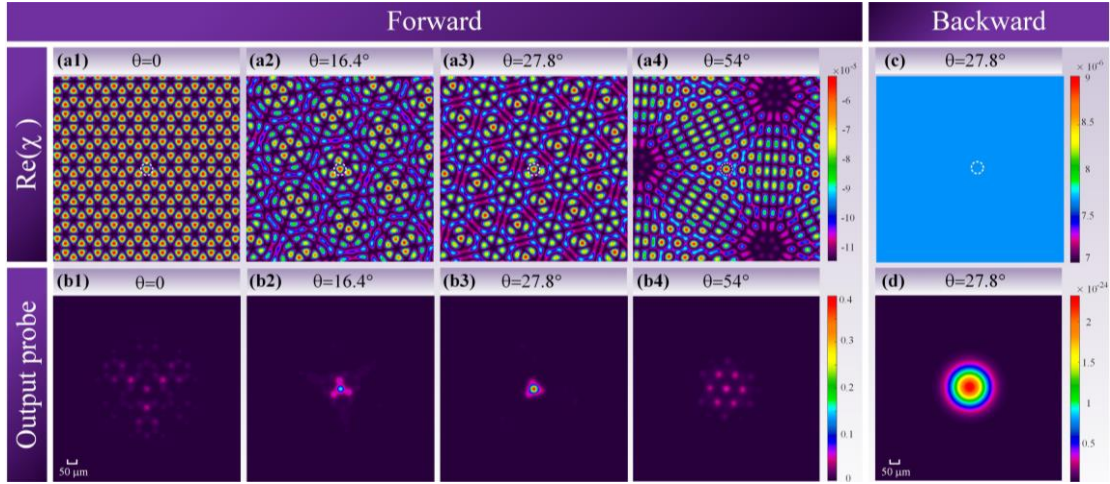

Fig. 2. The simulations of nonreciprocal transverse localization. (a1)-(a4) The real part of the susceptibility of moiré lattices with different twist angles of θ=0, 16.4°, 27.8° and 54.0° in the forward case. The dash circles mark the injecting position of the probe beam. (b1)-(b4) Simulated output patterns of the forward probe. (c) and (d) give the real part of the susceptibility and output pattern in the backward case with θ=27.8°, respectively. The amplitude of the periodic Rabi frequency of coupling filed is 165 MHz, and two-photon detuning is $\Delta_p - \Delta_c = 8$ MHz with $\Delta_p = 68$ MHz.

For the forward case with Doppler-free condition satisfied, the coupling field generates EIT effectively and induces a spatial periodic susceptibility distribution inside the atomic vapors. The resulted linear susceptibility is described as [32]:

$$\chi = \frac{iN|\mu_{31}|^2}{\hbar\varepsilon_0} \times \frac{1}{\left(\Gamma_{31} + i\Delta_p + [\Omega_c(x,y)]^2 / [\Gamma_{32} + i(\Delta_p - \Delta_c)]\right)}. \quad (1)$$

In the above expression, $\mu_{mn}$ and $\Gamma_{mn}$ are the respective dipole moment and decay rate between states $|m\rangle$ and $|n\rangle$ (m, n=1, 2, 3); N is the atomic density; $\varepsilon_0$ is the vacuum dielectric constant; $\Omega_c(x,y)=\mu_{mn}E_c(x,y)/\hbar$ is the spatially distributed Rabi frequency of the coupling field, with $E_c$ being the amplitude of the electric field.

According to Eq. (1), the simulated real parts of the susceptibility of the moiré photonic lattices obtained by superposing two honeycomb substructures with different twist angles θ are shown in Figs. 2(a1)-2(a4). When θ is 0, the susceptibility exhibits a honeycomb distribution. As θ increases to 16.4°,

27.8° and 54.0°, the profiles of the susceptibility are clearly distinctive from each other, but the translational symmetry is present in three cases with primitive cells of very different sizes, advocating the establishment of moiré photonic lattices. Here, the imaginary parts are nearly 2 orders of magnitude smaller than the real parts (Fig. S2 in Supplementary Materials), and can cause only a very weak absorption of the probe.

The propagation dynamics of the light inside the formed photonic lattices is checked by employing the Schrödinger-like paraxial equation [48]:

$$i\frac{\partial \psi(x,y,z)}{\partial z} = -\frac{1}{2k_0}(\frac{\partial^2}{\partial x^2}+\frac{\partial^2}{\partial y^2})\psi - \frac{k_0}{n_0}\Delta n(x,y)\psi, \qquad (2)$$

where $\psi$ is the envelope of the incident probe field $E_p$; $z$ is the propagation distance; $k_0=2\pi n_0/\lambda_p$ and $\lambda_p$ are the probe-field wavenumber and wavelength, respectively, with $n_0 \approx 1$ being the background refractive index; $\Delta n \approx \chi/2$ is the modulation of the refractive index arising from the coupling field.

The simulated output patterns of a narrow Gaussian forward-propagating probe in the transverse $x$-$y$ plane corresponding to different rotating θ of the coupling beams are given in Figs. 2(b1)-2(b4). The setting of θ=0 shows the expected honeycomb profile. The beam expansion is due to the evanescent coupling between neighboring waveguide channels, allowing a direct mapping to the honeycomb lattice of graphene. For the moiré lattices of θ≠0, only in the case of θ=27.8° a localized wavepacket is clearly seen, while the other twist angles lead to expanding patterns, with intensity modulation corresponding to the profile of Re($\chi$).

Figures 2(c) shows the simulated real part of refractive index in the backward case with θ=27.8° according to Eq. (S1) in Supplementary Material. Actually, Eq. (S1) has the similar form as Eq. (1), but with the term $\Delta_p-\Delta_c$ in Eq. (1) replaced by $\Delta_p-\Delta_c-2k_pv$. Obviously, when the two-photon resonance $\Delta_p-\Delta_c=0$ in the forward case is satisfied, term $\Delta_p-\Delta_c-2k_pv$ is hundreds of megahertz and far away from the resonance. As a consequence, the backward $\chi$ possesses a uniform distribution, which is independent on θ, since the Doppler frequency shift suppresses the interaction with the coupling field. Also, the backward imaginary part [see Fig. S2(e)], due to the absence of EIT, is about three orders of magnitude larger than that of the forward case. Therefore, the backward output based on Eq. (2) is much weaker, as shown in Fig. 2(d). Also, compared to the localized state in Fig. 2(b3), the backward transmitted pattern in Fig. 2(d) is much broader. In this sense, the nonreciprocal behaviors in both the transverse profile and the transmitted intensity occur in the designed atomic system.

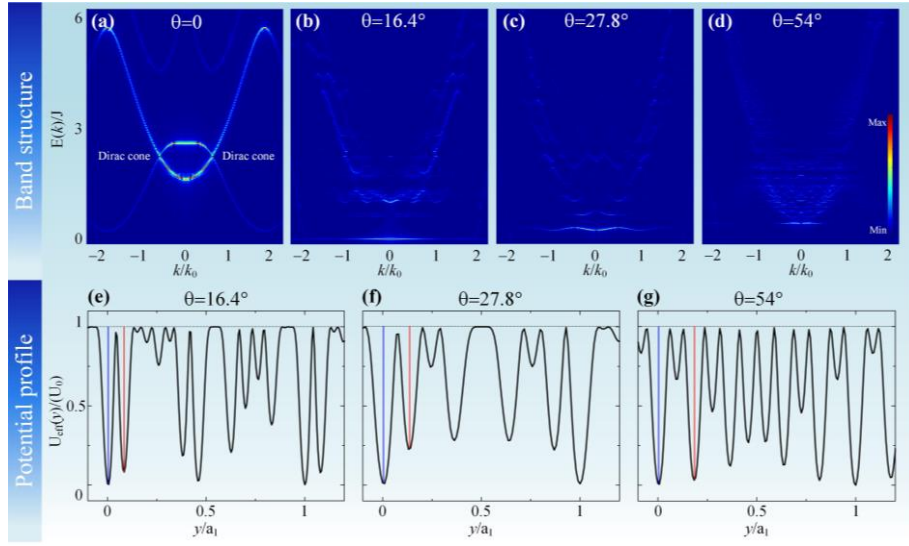

Fig. 3. Analyzations of the bands and potential profiles. (a)-(d) The energy bands of the moiré photonic lattices with different twist angles. (e)-(g) The profiles of potential inside different lattices according to Fig. 2(a). The red and blue vertical lines mark the deepest and the second deepest potentials, respectively. $U_{eff}(y)$ is the effective potential, with $U_0$ being peak potential of individual moiré lattice. $y/a_1$ is the normalized coordinate with $a_1$ being the minimum distance of two sites with the deepest potential in corresponding moiré lattices.

In general, the dispersionless (flat) band in photonic moiré structures can lead to localization, but the factors that can produce localization in periodic photonic structures are often much more complicated and go far beyond the flat-band effect. Figures 3(a)-3(d) show the corresponding energy bands, obtained by Fourier-transforming the solution of Eq. (2) with a narrow Gaussian input. The Dirac cones, typical for the honeycomb lattices, can be seen in Fig. 3(a) with $\theta=0$. By increasing $\theta$ to 16.4°, 27.8° and 54.0°, the moiré lattices in Figs. 3(b)-3(d) all demonstrate the formation of flat bands [the width of the narrowest band is $\delta E=6$ J for a honeycomb lattice, $\delta E \approx 0.13$ J in Fig. 3(b), $\delta E \approx 0.4$ J in Fig. 3(c) and $\delta E \approx 0.55$ J in Fig. 3(d)] from the original s-band of the honeycomb lattice. In principle, bands with low dispersion could effectively suppress the spreading of incident probe beams, if the beams were exciting only these flat bands. However, according to the simulations in Fig. 2(b), only $\theta=27.8°$ brings about a strong localization effect.

This unexpected result is understood by analyzing the profiles of potential inside different moiré cells, as given in Figs. 3(e) to 3(g), rather than only the band structures. On the one hand, the simulated moiré bands are not completely flat and still exhibit slight non-zero dispersions, and the probe beam is exciting multiple bands, which will unavoidably lead to the broadening of beams. On the other hand, the moiré lattice obtained by interference of the coupling beams via EIT exhibits an important peculiarity: contrary to other moiré lattices studied in electronics and photonics [38], the sites of our lattices exhibit different potential depths. During the probe beam propagation, tunneling is only possible between sites (waveguide channels) which have a comparable depth of potential. Our probe beam is always focused on the site with the deepest potential (the largest susceptibility), see Fig. 2(a). For θ=27.8°, the potential of the second deepest site is only 79% of that in the deepest one. Whereas in θ=16.4° (θ=54.0°), the second deepest potential in the vicinity of the excited site is 91% (97%) of the deepest one, respectively. Smaller difference of the susceptibility between two sites makes tunneling much easier.

By analyzing the difference of the susceptibility between two adjacent sites, one can understand the output results in Fig. 2(b) of different moiré lattices much better. With θ=16.4°, the probe beam is not fully localized and slightly couples into surrounding waveguides, due to the similar susceptibilities of the incident site and circumambient sites. The further enlargement of the difference in θ=28.7° suppresses the coupling of the light between adjacent channels: the transverse localization occurs. A very minor portion of the probe expands outside the incident site. Compared to that in θ=0 with large dispersion, the localization in θ=28.7° improves the output peak intensity ten times. The transmitted pattern in θ=54.0° exhibits the strongest broadening, because the involved sites have almost the same potential. It's worth noting that the suppression on the broadening of wave packets from the formed flat bands in all cases is verified by the fact that the sizes of the output patterns in θ≠0 are smaller than in the case of θ=0 with dispersive bands.

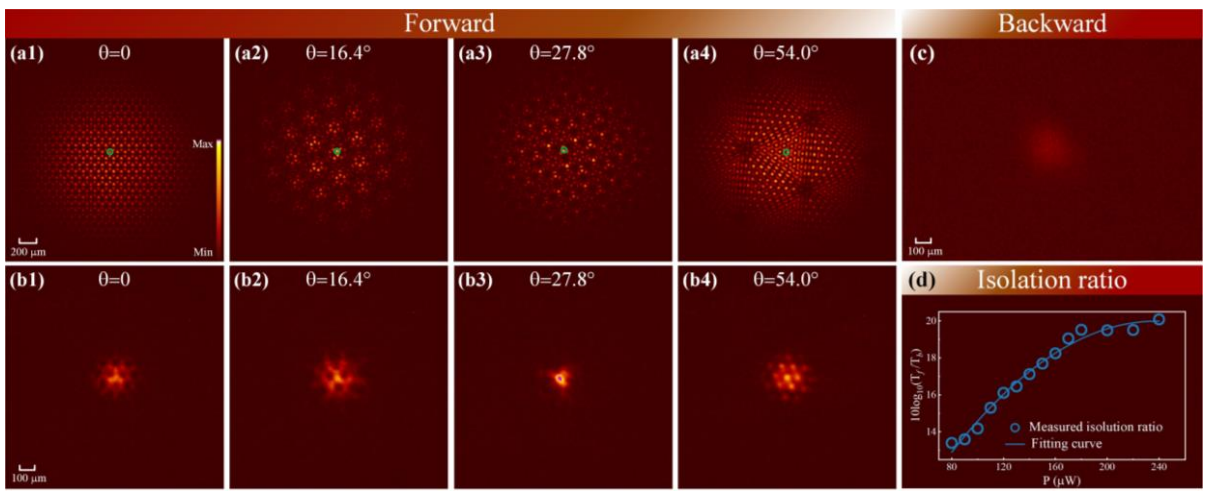

Fig. 4. Experimental observation of nonreciprocal transverse localization. (a) The obtained intensity patterns of coupling field superposed of two honeycomb substructures with different twist angles of 0, 16.4°, 27.8° and 54.0°. The green circles indicate the incident position of the probe beam. (b) The corresponding output patterns of the forward probe beam. (c) The transmission of the backward probe beam at θ=27.8°. (d) The dependence of the isolation ratio on the input power $P$ when both probe beams with the same parameters are present simultaneously (experimental measurements – circles). The curve is a guide to the eye.

Figure 4(a) shows the experimentally established spatially periodic coupling field composed of two honeycomb substructures with different twist angles. As shown in Fig. 4(a1) with θ=0, the resulted compound pattern still shows a honeycomb distribution. The coupling fields with different moiré patterns by changing θ are presented in Figs. 4(a2)-4(a4), which support the predictions in Fig. 2(a) well. The waveguide channels are optically induced at the positions with the largest intensity of the structured coupling field.

When the forward probe beam is focused onto one site (marked by the green circle) of the coupling field, the output patterns are obtained as shown in Fig. 4(b). Here the size of the probe wave packet at the input surface of the vapor cell is comparable to that of one lattice site. Figure 4(b1) shows a clear honeycomb profile under the condition of θ=0 and confirms the formation of a honeycomb photonic lattice inside the cell. By setting θ=16.4°, the moiré pattern in Fig. 4(a2) shows that the primitive cell contains several honeycomb units and is quite different from that at θ=0. As a result, the output probe in Fig. 4(b2) is broadened and demonstrates a modulation corresponding to the moiré lattice. With the angle further increased to θ=27.8°, the output probe in Fig. 4(b3) shows a narrow wave packet, whose size is close to that of the incident probe beam. This is the transverse localization caused by the suppressed band dispersion and the relatively large difference of potential between the incident and surrounding channels. For the result of Fig. 4(b4), the expansion of the probe beam is recovered and its output pattern shows a clear hexagonal modulation, agreeing well with the moiré pattern of θ=54°. All the experimental observations coincide with the simulated Fig. 2(b).

Figure 4(c) shows the output pattern of the backward probe beam for θ=27.8°, displaying a broad Gaussian profile with low intensity due to the unavoidable resonant absorption. The backward pattern does not exhibit obvious change with modifying θ, because the lack of EIT makes the backward probe field immune to the modulation of coupling field. For the case of θ=27.8°, the forward output demonstrates a localized state with strong transmission, while the backward output gives a spreading and weak pattern. These observations constitute the effect of nonreciprocal transverse localization. The localization factor is ~2.85, which is defined as the ratio of FWHM of the output probe in the backward and forward cases.

Figure 4(d) shows the dependence of isolation ratio on the input power $P$ when both probe beams are oppositely but collinearly sent into the medium simultaneously under $\theta=27.8°$. The transmitted intensities of the forward and backward probe are defined as $T_f$ and $T_b$, and the isolation ratio is obtained as $10\log_{10}(T_f/T_b)$. Here $T_f$ and $T_b$ are acquired by integrating the intensity in the capturing region of the cameras. The experimentally transmitted patterns of the forward and backward probe beams are shown in Fig. S3 and S4 in Supplementary Material, respectively.

By increasing the probe power $P$ from 80 μW to 240 μW, the isolation ratio grows from 13 dB to 20.1 dB. For the forward case with EIT, the enhancement of the input probe intensity can lead to a stronger $T_f$, since the absorptive coefficient is determined by the imaginary part of the susceptibility, which does not rely on the probe intensity according to Eq. (1). For the backward case, the probe beam experiences a very strong loss due to the relatively large atomic density and absence of EIT from the coupling beams, and the transmission $T_b$ approaches nearly 0. Even when the probe power is increased to 240 μW, the backward transmission is still very weak. This is probably due to the non-linear susceptibility (both real and imaginary), which is also different in the forward and backward cases. As a result, the isolation ratio can increase with the probe power. The measurement with two probe beams turned on together advocates that the demonstrated nonreciprocal localization can counteract dynamic reciprocity.

In summary, by exploiting a SLM to build superposed patterns consisting of two honeycomb substructures, moiré photonic lattices under different twist angles are effectively established inside atomic vapors with the presence of EIT. In particular, the moiré photonic lattice with a selective twist angle can give rise to desired transverse localization for the probe beam. The combination of the intrinsic Doppler effect from thermal motion of atoms (which is usually considered to be detrimental in the formation of various coherent effects) and the established transverse localization provides an effective route to realize nonreciprocal localization of photons. What's more, the isolation ratio of the current regime can be promisingly improved by introducing gain with an additional pump beam [54, 55]. The nonreciprocal transverse localization that we have demonstrated can be used in magnetic-free nonreciprocal optical devices bypassing dispersive effects for the incident beams and promisingly improving the matching between the output beams and subsequent optical components for signal processing, with such a wide range of applications as one-way optical imaging or sensing, directional optical communications, and nonreciprocal traps.

## Method

**Experimental settings.** Both the wavelengths of the probe and coupling fields are ~795.0 nm, and their frequency difference is about $2\pi\times3.03$ GHz. The probe beam is from an external-cavity diode laser, while the coupling beam is derived from a semi-conductive tapered amplifier seeding with a beam from the other external-cavity diode laser. The power of the coupling field with a moiré intensity pattern is about 40 mW. The adopted phase-type liquid crystal SLM has a resolution of 1920×1152. The Rb cell with a length of 5 cm is heated to 120°C by a home-made temperature controller.

## Data availability

The data that supports the results within this paper are available from the corresponding authors upon reasonable request.

## Authors contribution

Z. Z. conceived the original idea and supervised the project. D. S. and H. J. co-supervised the project. S. L. and C. C. L. conducted the experiment with the help from Z. Z. and Y. Z. S. L., C. C. L., W. Y., Z. L., G. M. and D. S. provided the theoretical support. D. S., H. J., G. M., C. B. L., Y. Z. and Z. Z. contributed to interpretation of experimental results. Z. Z., D. S. and S. L. wrote the manuscript with inputs from all authors.

## Acknowledgements

This work was supported by National Natural Science Foundation of China (No. 62475209), Qinchuangyuan "Scientist+Engineer" Team Construction of Shaanxi Province (2024QCY-KXJ-178), and by the European Union's Horizon 2020 program, through a FET Open research and innovation action under the grant agreements No. 964770 (TopoLight). Additional support was provided by the ANR Labex GaNext (ANR-11-LABX-0014), the ANR program "Investissements d'Avenir" through the IDEX-ISITE initiative 16-IDEX-0001 (CAP 20-25), the ANR project MoirePlusPlus (ANR-23-CE09-0033), and the ANR project "NEWAVE" (ANR-21-CE24-0019).

## Competing interests

The authors declare no competing interests.

## Additional information

**Supplementary information** The online version contains supplementary material available at https://XX.

# Supplemental Materials for
# Observation of nonreciprocal transverse localization of light

Shun Liang[1], Changchang Li[1], Wenqing Yu[1], Zhenzhi Liu[1], Changbiao Li[1], Yanpeng Zhang[1],

Guillaume Malpuech[2], Dmitry Solnyshkov[2,3,†], Hui Jing[4,*], Zhaoyang Zhang[1,‡]

[1]Key Laboratory for Physical Electronics and Devices of the Ministry of Education & Shaanxi Key Lab of Information Photonic Technique, School of Electronic and Information Engineering, Faculty of Electronic and Information Engineering, Xi'an Jiaotong University, Xi'an, 710049, China

[3]Institut Pascal, PHOTON-N2, Université Clermont Auvergne, CNRS, Clermont INP, F-63000 Clermont-Ferrand, France

[4]Institut Universitaire de France (IUF), 75231 Paris, France

[2]Key Laboratory of Low-Dimensional Quantum Structures and Quantum Control of Ministry of Education, Department of Physics and Synergetic Innovation Center for Quantum Effects and Applications, Hunan Normal University, Changsha 410081, China

Emails: †dmitry.solnyshkov@uca.fr, *jinghui@hunnu.edu.cn, ‡zhyzhang@xjtu.edu.cn

## I. Experimental Setup

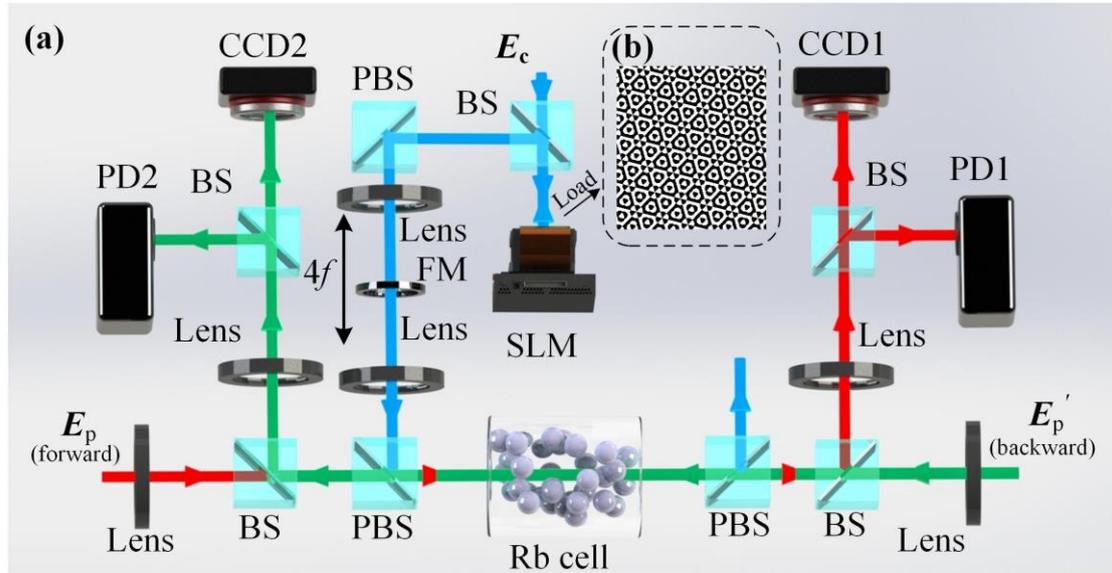

Fig S1. Experimental setup. A SLM is employed to modulate the vertically polarized coupling field $E_c$ into a moiré pattern consisting of two honeycomb substructures. Two horizontally polarized probe beams $E_p$ (forward case) and $E_p'$ (backward case) are sent into the vapor cell in opposite directions, and their output patterns propagating through the atomic medium are imaged onto two cameras. Two photodiode detectors are placed to monitor the transmitted probe spectra in forward and backward cases. PBS: polarization beam splitter; BS: beam splitter; FM: Fourier mask; CCD: charge-coupled device camera; PD: photodiode detector. (b) The calculated holography loaded onto the SLM for generating the moiré pattern with a twist angle of 27.8°.

Figure S1(a) depicts the experimental setup. The coupling field $E_c$ with a Gaussian profile is incident onto the spatial light modulator (SLM) loaded a phase holography

[Fig. S1(b) shows the phase holography with θ=27.8° as an example], the reflected part experiencing phase modulation will exhibit discretized intensity distribution. A *4f* optical system composed of two lenses is introduced to produce pseudo-nondiffracting optical beam [1, 2], which ensures that the distribution of the obtained coupling field with a moiré pattern keeps almost unchanged during the travelling inside the Rb cell. A Fourier mask is inserted into the *4f* system to filter out unwanted high-order diffraction components. When $E_c$ with the desired profile propagates through the Rb cell, two oppositely injecting probe fields $E_p$ (forward) and $E_p'$ (backward) are focused onto the same channel of the discrete coupling field. Then, the output $E_p$ ($E_p'$) is imaged to the CCD1 (CCD2) by individual imaging lens. In addition, the beam splitter in front of CCD1 (CCD2) splits $E_p$ ($E_p'$) so that a part of the beam can be received by the PD1 (PD2) to monitor the transmitted spectrum.

## II. Susceptibility

The main text provides the Eq. (1) for the susceptibility when the probe field and the coupling field propagate in the same direction, by which the Doppler shift is cancelled. When the backward probe beam $E_p'$ and $E_c$ are incident into the Rb cell oppositely, the EIT condition is destroyed by the Doppler shift, and the susceptibility experienced by $E_p'$ can be described as:

$$\chi^{(1)}dv = \frac{iN|\mu_{31}|^2}{\hbar\varepsilon_0} \times \frac{1}{\left(\Gamma_{31} + i(\Delta_p - k_p v) + \frac{[\Omega_c(x,y)]^2}{\Gamma_{32} + i(\Delta_p - \Delta_c) - i(k_p v + k_c v)}\right)} \times \frac{e^{-v^2/u^2}}{u\sqrt{\pi}} dv, \quad (S1)$$

where $k_p$ ($k_c$) is the wave number of the probe (coupling) beam, $N(v) = e^{-v^2/u^2}/u\sqrt{\pi}$ is the Maxwell-Boltzmann distribution of velocity, with *u* being the most probable velocity, and velocity *v* being the atomic velocity.

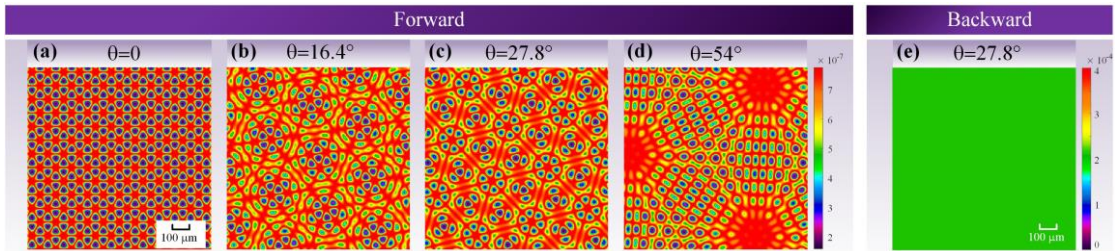

Fig. S2. The theoretical imaginary parts of susceptibilities in the forward case with different twist angles (a) 0°, (b) 16°, (c) 27.8° and (d) 54°. (e) The theoretical imaginary parts of the backward case with the twist angle being 27.8°.

Figure S2(a-d) shows the distribution of the imaginary parts of the susceptibility in the forward case according to Eq. (1) in main text. Compared to the real part of the susceptibility in Fig. 2(a) in main text, the imaginary part is two orders of magnitude smaller than the real part. This result indicates that the propagation of the probe beam inside the moiré lattices is dominated by the real part of the susceptibility. Figure S2(e) shows the imaginary part of the susceptibility in the backward case with a rotation angle of 27.8°. The larger backward imaginary part means a much stronger absorption.

## II. Experimentally Output Probe Patterns

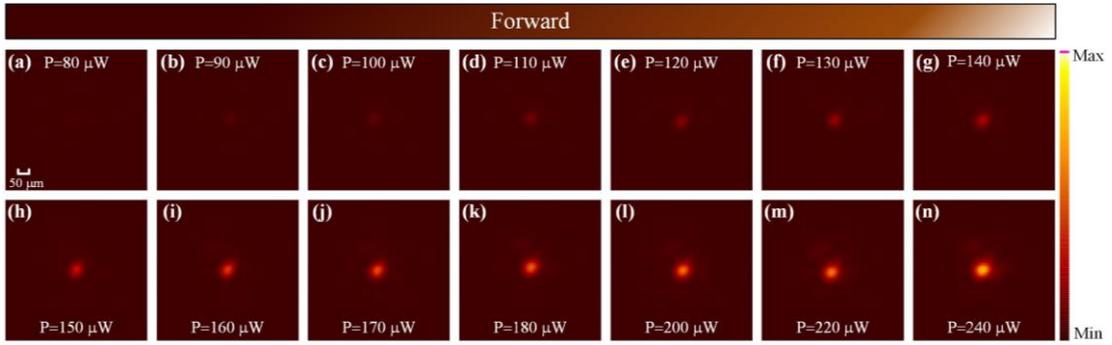

Fig. S3 The experimental evolutions of the output forward probe pattern with its power increased from 100 μW to 240 μW. The twist angle is 27.8°.

Figures S3 and S4 shows the output patterns of the forward and backward probe fields by varying their powers synchronously, when they are oppositely sent into the medium simultaneously under the case of θ=27.8°. According to Fig. S3, one can see that the output intensity obviously increases with the probe power growing from 100 μW to 240 μW. Since the adopted camera can respond only in a limited range of the incident power, an attenuator (with a ratio of 19.5 dB in intensity) is placed before the camera for capturing the forward probe. Such an attenuator makes the camera work below the threshold for saturation.

Whereas for the case of the power below 160 μW in Fig. S4, the backward output patterns are so weak that they are difficult to be recognized by eyes, due to the break of the EIT condition. Moreover, the absence of EIT suppresses the effect of the coupling field on the probe beam. No obvious modulation can be seen in the spatial profile of the output probe, which remains a broadened Gaussian.

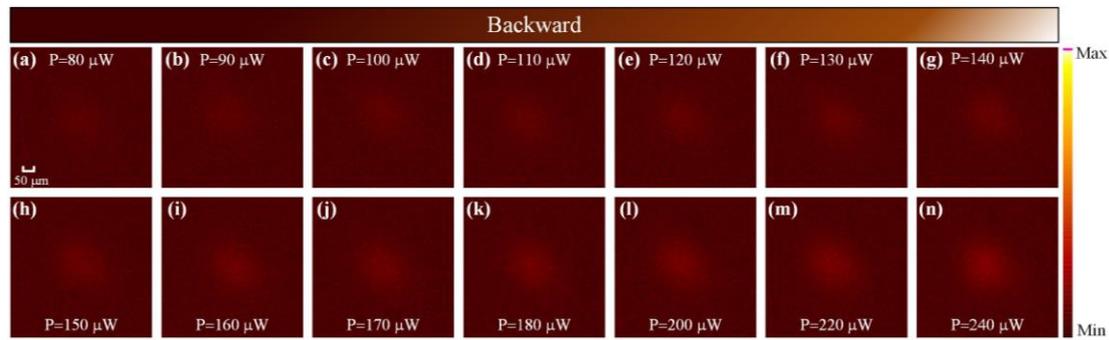

Fig. S4 The backward output probe patterns corresponding to Fig. S3.